\documentclass[twocolumn,showpacs,aps]{revtex4}

\usepackage{graphicx}
\usepackage{latexsym}
\usepackage{amssymb}
\usepackage{color}
\usepackage[center]{subfigure}

\begin{document}

\newcommand{\bq}{\begin{equation}}
\newcommand{\eq}{\end{equation}}
\newcommand{\bqn}{\begin{eqnarray}}
\newcommand{\eqn}{\end{eqnarray}}
\newcommand{\nb}{\nonumber}
\newcommand{\lb}{\label}

\title{Vaidya Solutions in General Covariant Ho\v{r}ava-Lifshitz Gravity 
without Projectability: Infrared Limit}

\author{O. Goldoni $^{1}$}
\email{otaviosama@gmail.com}
\author{M.F.A. da Silva $^{1}$}
\email{mfasnic@gmail.com}
\author{G. Pinheiro $^{1}$}
\email{gpinheiro.fisica@gmail.com}
\author{R. Chan $^{2}$}
\email{chan@on.br}
\affiliation{
$^{1}$ Departamento de F\'{\i}sica Te\'orica,
Instituto de F\'{\i}sica, Universidade do Estado do Rio de Janeiro,
Rua S\~ao Francisco Xavier 524, Maracan\~a
20550-900, Rio de Janeiro, RJ, Brasil.\\
$^{2}$ Coordena\c{c}\~ao de Astronomia e Astrof\'{\i}sica, 
Observat\'orio Nacional, Rua General Jos\'e Cristino, 77, S\~ao Crist\'ov\~ao  
20921-400, Rio de Janeiro, RJ, Brazil.
}

\date{\today}

\begin{abstract}
In this paper, we have studied nonstationary radiative spherically symmetric 
spacetime, in general covariant theory ($U(1)$ extension) of 
Ho\v{r}ava-Lifshitz 
gravity without the projectability condition and in the infrared limit. 
The Newtonian prepotential $\varphi$ was assumed null.
We have shown that there is not the analogue of the Vaidya's solution in the 
Ho\v{r}ava-Lifshitz Theory (HLT), as we know in the General Relativity Theory 
(GRT).  Therefore,
we conclude that the gauge field $A$ should interact with the null radiation
field of the Vaidya's spacetime in the HLT. 
\end{abstract}

\pacs{04.50.Kd; 98.80.-k; 98.80.Bp}

\maketitle

\section{Introduction}

One of the biggest problem of the GRT lies on the difficult of its
quantization, since it is a non-renormalizable theory. However,
Ho\v{r}ava \cite{Horava} has proposed a benchmark in renormalizable quantum 
gravity
theory which has attracted a great interest. The theory was inspired by the 
Lifshitz scalar \cite{Lifshitz} and has often been called Ho\v{r}ava-–Lifshitz 
theory gravity.  
He has formulated a theory of quantum gravity, whose scaling at 
short distances exhibits a strong anisotropy between space and time 
\cite{Horava},
\bq
\lb{1.1}
{\bf x} \rightarrow b^{-1} {\bf x}, \;\;\;  t \rightarrow b^{-z} t.
\eq
In order for the theory to be power-counting renormalizable, in 
$(3+1)$-dimensions the critical exponent $z$  needs to be $z \ge 3$ 
\cite{Horava,Visser}.  Since the literature about the theory is very extensive
we suggest the reader the references \cite{BPS}-\cite{GMW}.

In order to solve several problems in the HLT, Wang and collaborators have 
proposed a model without the projectability 
condition, but assuming that:  (a) the detailed balance condition is softly 
broken; and (b) the symmetry of theory is enlarged to included a local $U(1)$ 
symmetry \cite{ZWWS,ZSWW}. The enlarged symmetry was first introduced
by Ho\v{r}ava and Melby-Thompson (HMT) in  the case with the projectability 
condition and $\lambda = 1$ \cite{HMT}, and was soon generalized to the case 
with any $\lambda$ \cite{Silva}, where $\lambda$ is a coupling constant, which 
characterizes the deviation from GRT in the infrared limit
\cite{HMT,WW,Silva,HW,LWWZ,Kluson2}. 

In this paper, we will analyze if the Vaidya's spacetime can be described
as a null radiation fluid in the general
covariant HLT of gravity without the projectability condition \cite{ZWWS,ZSWW}.
In Section II we present a brief introduction to the HLT. In Section III
we show the Vaidya's spacetime, expressed in ADM decomposition\cite{ADM}.  
In Section IV
we present the HLT equations for the infrared limit and their possible
solutions. In Section V we analyze all the possible solutions for the
HLT field equations. In Section VI we discuss the results. Finally, in  
Appendix A we present all the equations of HLT without projectability.

\section{General Covariant Ho\v{r}ava-Lifshitz Gravity without Projectability}

In this section, we shall give a very brief introduction to the general 
covariant HLT gravity without the projectability condition. For detail, we 
refer readers to \cite{ZWWS,ZSWW}.
The total action of the theory can be written as,
\bqn 
\lb{TA}
S &=& \zeta^2\int dt d^{3}x  \sqrt{g}N \Big({\cal{L}}_{K} -
{\cal{L}}_{{V}} +  {\cal{L}}_{{A}}+ {\cal{L}}_{{\varphi}}  + \frac{1}{\zeta^2} {\cal{L}}_{M}\Big), \nb\\
\eqn
where $g={\rm det}(g_{ij})$, and
\bqn \lb{2.5}
{\cal{L}}_{K} &=& K_{ij}K^{ij} -   \lambda K^{2},\nb\\
{\cal{L}}_{V} &=&  -  \Big(\beta_0  a_{i}a^{i}- \gamma_1R\Big),\nb\\
{\cal{L}}_{A} &=&\frac{A}{N}\Big(2\Lambda_{g} - R\Big), \nb\\
{\cal{L}}_{\varphi} &=&  \varphi{\cal{G}}^{ij}\big(2K_{ij}+\nabla_i\nabla_j\varphi+a_i\nabla_j\varphi\big)\nb\\
& & +(1-\lambda)\Big[\big(\Delta\varphi+a_i\nabla^i\varphi\big)^2  
+2\big(\Delta\varphi+a_i\nabla^i\varphi\big)K\Big]\nb\\
& & +\frac{1}{3}\hat{\cal G}^{ijlk}\Big[4\left(\nabla_{i}\nabla_{j}\varphi\right) a_{(k}\nabla_{l)}\varphi \nb\\
&&  ~~ + 5 \left(a_{(i}\nabla_{j)}\varphi\right) a_{(k}\nabla_{l)}\varphi + 2 \left(\nabla_{(i}\varphi\right)a_{j)(k}\nabla_{l)}\varphi \nb\\
&&
~~ 
+ 6K_{ij} a_{(l}\nabla_{k)}\varphi 
\Big],
\eqn
where $A$ and $\varphi$ are the the gauge field and the Newtonian prepotential,
respectively \cite{Lin2013}. 
Here $\Delta \equiv g^{ij}\nabla_{i}\nabla_{j}$, $\Lambda_{g}$ is a coupling 
constant, and all the coefficients, $ \beta_{n}$ and $\gamma_{n}$, are
dimensionless and arbitrary, except for the ones of the sixth-order derivative 
terms, $\gamma_{5}$ and $\beta_{8}$, which must satisfy the conditions, 
\bq
\lb{2.8a}
\gamma_{5} > 0, \;\;\; \beta_{8} <  0,
\eq
in order to the theory to be unitary in the UV. The Ricci and Riemann tensors 
$R_{ij}$ and $R^{i}_{\;\; jkl}$  all refer to the 3-metric $g_{ij}$, with 
$R_{ij} = R^{k}_{\;\;ikj}$ and
\bqn \lb{2.6}
K_{ij} &\equiv& \frac{1}{2N}\left(- \dot{g}_{ij} + \nabla_{i}N_{j} +
\nabla_{j}N_{i}\right),\nb\\
{\cal{G}}_{ij} &\equiv& R_{ij} - \frac{1}{2}g_{ij}R + \Lambda_{g} g_{ij}.
\eqn
${\cal{L}}_M$ is the Lagrangian of matter fields.  To be consistent with 
observations in the infrared limit, we assume that
\bq
\lb{2.8}
\zeta^{2} = \frac{1}{16\pi G},\;\;\;  \gamma_{1} = -1,
\eq
where $G$ denotes the Newtonian constant, and
\bq
\lb{2.8b}
\Lambda \equiv \frac{1}{2} \zeta^{2}\gamma_{0},
\eq
is the cosmological constant. $C_{ij}$ denotes the Cotton tensor, defined by
\bq
\lb{1.12}
C^{ij} = \frac{ {{e}}^{ikl}}{\sqrt{g}} \nabla_{k}\Big(R^{j}_{l} - \frac{1}{4}R\delta^{j}_{l}\Big),
\eq
with  $e^{123} = 1$.  Using the Bianchi identities, one can show that 
$C_{ij}C^{ij}$ can be written in terms of the five independent sixth-order 
derivative terms in the form
\bqn
\lb{1.13}
C_{ij}C^{ij}  &=& \frac{1}{2}R^{3} - \frac{5}{2}RR_{ij}R^{ij} + 3 R^{i}_{j}R^{j}_{k}R^{k}_{i}  +\frac{3}{8}R\Delta R\nb\\
& &  +
\left(\nabla_{i}R_{jk}\right) \left(\nabla^{i}R^{jk}\right) +   \nabla_{k} G^{k},
\eqn
where
\lb{1.14}
\bqn
G^{k}=\frac{1}{2} R^{jk} \nabla_j R - R_{ij} \nabla^j R^{ik}-\frac{3}{8}R\nabla^k R.
\eqn

Variations of the total action (\ref{TA}) with respect to $N, \; N^{i}, \; A, 
\; \varphi$ and $g_{ij}$ yield, respectively, the Hamiltonian, momentum, $A$-, 
and $\varphi$-constraints, and dynamical equations, which are given explicitly 
in \cite{ZSWW}. For the sake of reader's convenience, we include them in 
Appendix A. 
  
In addition, assuming the translation symmetry of the action, 
one obtains the conservation laws of energy and momentum \cite{ZSWW}, which 
are also given in Appendix A.

\section{Vaidya's Spacetime}

The Arnowitt-Deser-Misner (ADM) form is given by \cite{ADM},
\bqn
ds^{2} &=& - N^{2}dt^{2} + g_{ij}\left(dx^{i} + N^{i}dt\right)
\left(dx^{j} + N^{j}dt\right), \nb\\
& & ~~~~~~~~~~~~~~~~~~~~~~~~~~~~~~  (i, \; j = 1, 2, 3).
\eqn

Hereinafter, the Newtonian prepotential $\varphi$ is assumed null and $G=c=1$.

The Vaidya's spacetime with an ingoing null dust usually written in the form 
\cite{BS09},
\bq
\lb{3.1}
ds^2 = - \left(1 - \frac{2m(v)}{r}\right) dv^2 + 2dvdr + r^2d\Omega^2,
\eq
where $d\Omega^2 \equiv d\theta^2 + \sin^2\theta d\phi^2$, and the 
corresponding energy-momentum tensor is given by
\bq
\lb{3.2}
T_{\mu\nu} = \rho(v,r)l_{\mu}l_{\nu},
\eq
with
\bq
\lb{3.3}
\rho =  \frac{2}{r^2}\frac{dm}{dv},\;\;\; l_{\mu} = - \delta_{\mu}^{v}.
\eq

Introducing a time-like coordinate $t$ via the relation, $v = 2(t + r)$, the 
metric (\ref{3.1}) can be cast in the form,
\bqn
\lb{metric}
ds^2 &=& - \frac{r}{M}dt^2 + \frac{4M}{r}\left[dr + \left(1 - \frac{r}{2M}\right)dt\right]^2 \nb\\
&& + r^2d\Omega^2,
\eqn
where 
\bq
\lb{3.4}
M \equiv M(V) = 2m(v),\;\;\; V \equiv t + r. 
\eq
From equation (\ref{metric}), we immediately obtain 
\bqn
\lb{3.5a}
N &=& \sqrt{\frac{r}{M}},\;\;\; N^{i} = \left(1 - \frac{r}{2M}\right)\delta^{i}_{r},\nb\\
g_{rr} &=& \frac{4M}{r},\;\;\; g_{\theta \theta} = r^2, \;\;\;  g_{\phi \phi} = r^2\sin^2\theta,
\eqn
and
\bqn
\lb{3.5b}
N_{i} &\equiv& g_{ij}N^{j} =   -2 \left(1 - \frac{2M}{r}\right)\delta_{i}^{r},\nb\\
g^{rr} &=& \frac{r}{4M},\;\;\; g^{\theta\theta} = \frac{1}{r^2}, \;\;\;  g^{\phi\phi} = \frac{1}{r^2\sin^2\theta},\\
\rho &=& \frac{M^{*}}{2r^2},\;\;\; l_{\mu} = - 2\left(\delta^{t}_{\mu} + \delta^{r}_{\mu}\right),
\eqn
where ${M}^{*}  \equiv dM/dV$.  

Since $M=M(V)$, introducing another independent variable, $U=t-r$, we can 
find that 
\bq
M'=\dot M= \frac{1}{2}M^{*},
\lb{M*}
\eq
since $dM(V)/dU=0$.

Then, we find that the non null metric components are
\bqn
\lb{3.5c}
{}^{(4)}g_{tt} &=& -\left(N^2 - N_{i}N^{i}\right) = -\frac{4}{r}\left(M-r\right),\nb\\
{}^{(4)}g_{ti} &=& N_{i} = \frac{2}{r}\left(2M-r\right) \delta^r_i,\nb\\
{}^{(4)}g_{rr} &=& g_{rr}= \frac{4M}{r},\nb \\
{}^{(4)}g_{\theta\theta} &=& g_{\theta\theta}= r^2,\nb \\
{}^{(4)}g_{\phi\phi} &=& g_{\phi\phi}= r^2 \sin^2 \theta,\nb \\
{}^{(4)}g^{tt} &=& -\frac{1}{N^{2}} = -\frac{M}{r}  \nb\\
{}^{(4)}g^{ti} &=& \frac{N^{i}}{N^2} = \frac{M}{r}\left(1-\frac{r}{2M}\right) \delta^i_r,\nb\\
{}^{(4)}g^{rr} &=& 1-\frac{M}{r},\nb  \\
{}^{(4)}g^{\theta\theta} &=& \frac{1}{r^2},\nb  \\
{}^{(4)}g^{\phi\phi} &=& \frac{1}{r^2 \sin^2 \theta},\nb  \\
\eqn

\bqn
\lb{3.5d}
{}^{(4)}g^t_t &=& 1, \nb\\
{}^{(4)}g^r_r &=& 1, \nb \\
{}^{(4)}g^\theta_\theta &=& 1, \nb \\
{}^{(4)}g^\phi_\phi &=& 1. \nb \\
\eqn

For the projection tensor the non null components are
\bqn
\lb{3.5e}
{}^{(4)}h^r_t &=& 1-\frac{r}{2M}, \nb\\
{}^{(4)}h^r_r &=& 1, \nb \\
{}^{(4)}h^\theta_\theta &=& 1, \nb \\
{}^{(4)}h^\phi_\phi &=& 1. \nb \\
\eqn

Then, it can be shown that
\bqn
\lb{3.6}
n_{\mu} &=& N\delta^{t}_{\mu} = \sqrt{\frac{r}{M}},\nb\\
n^{\mu} &\equiv& {}^{(4)}g^{\mu\nu}n_{\nu} = -\frac{1}{N}\left(\delta^{\mu}_{t} - N^i\delta^{\mu}_{i}\right) \nb \\ 
&=& \sqrt{\frac{M}{r}}\left[-\delta^\mu_t+\left(1-\frac{r}{2M}\right) \delta^\mu_r\right], \\ 
h^\mu_\nu &=& ^{(4)}g^\mu_\nu + n^\mu n_{\nu}, \nb\\
h^r_t &=& 1-\frac{r}{2M}, \nb\\
h^r_r &=& 1, \nb\\
h^\theta_\theta &=& 1, \nb\\
h^\phi_\phi &=& 1, \\
J_i &=&  T_{\mu\nu}n^{\mu}h^{\nu}_{i}\nb \\
&=&\frac{1}{r^2}\sqrt{\frac{M}{r}}\frac{dM}{dV}\delta^r_i \nb \\
&=&\frac{1}{r^2}\sqrt{\frac{M}{r}}(\dot M + M')\delta^r_i,  \\
\lb{jdi}
\tau_{ij} &=&  T_{\mu\nu}h^{\mu}_{i}h^{\nu}_{j}\nb \\
&=& \frac{8M}{r^3}\frac{dM}{dV}\delta^r_i \delta^r_j\nb \\
&=&\frac{8M}{r^3}(\dot M + M')\delta^r_i\delta^r_j, \\ 
\eqn
where the prime and dot denotes the partial differentiation in relation to the
coordinate $r$ and $t$, respectively.

\section{Infrared Limit}

In the infrared limit we must have

\bq
J_t = -2 \rho.
\eq

Besides, hereinafter, we have assumed that $\lambda=1$.

Thus we have
\bqn
\lb{3.7}
K_{rr}&=&\frac{-2 M' M r-M' r^2-2 \dot M M r+2 M^2+M r}{\sqrt{r/M} M r^2}, \nb\\
K_{\theta\theta}&=&\frac{r(-2 M+r)}{2 \sqrt{r/M} M}, \nb\\
K_{\phi\phi}&=&\sin^2 \theta \frac{r (-2 M+r)}{2 \sqrt{r/M} M}, \nb \\
K &=& \frac{-2 M' M r-M' r  2-2 \dot M M r-6 M^2 +5 M r}{4 \sqrt{r/M} M^2 r}, \nb \\
R_{rr} &=& \frac{M' r-M}{M r^2}, \nb \\
R_{\theta\theta}&=&\frac{M' r^2+8 M^2-3 M r}{/8 M^2}, \nb \\
R_{\phi\phi}&=&\sin^2 \theta \frac{M' r^2+8 M^2-3 M r}{8 M^2}, 
\eqn
\bq
\lb{Riccscalar}
R = \frac{M' r^2+4 M^2-2 M r}{2 M^2 r^2}, 
\eq
and
\bqn
{\cal{L}}_K &=& \frac{1}{2 M^2 r^2} \times \nb \\ 
& &[-4 M'^2 M^2 + M' M r^2+4 \dot M M^2- 2 \dot M M r+\nb \\
& &4 M^2 - 2M r] 
\eqn
\bqn
{\cal{L}}_V&=&-\frac{1}{2 M^2 r^2} [M' r^2+4 M^2-2 M r]
\eqn
\bqn
F_V&=&\frac{\beta_0}{16 M^3 r}[-4 M'' M r^2+7 M'^2 r^2-14 M' M r+7 M^2]\nb \\
\eqn

From equation (\ref{mom}) we have
\bqn
H&=&{\cal{L}}_K + {\cal{L}}_V + F_V=8\pi J^t=\frac{1}{16 M^3 r^2}\times\nb \\
& &[-4 M'' \beta_0 M r^3+7 M'^2 \beta_0 r^3- 14 M'\beta_0 M r^2- \nb \\
& &32 M' M^3 +32 \dot M M^3 -16 \dot M M^2 r+7 \beta_0 M^2 r]
\eqn

\bq
J_r=\frac{\dot M}{ \sqrt{r/M} M r}
\lb{jdr}
\eq

\bq
J_A=\frac{1}{2 M^2 r^2}[M' r^2-4 \Lambda_g M^2 r^2+4 M^2-2 M r]
\eq

\bqn
J_{\varphi}&=&\frac{1}{16 \sqrt{r/M} M^4 r^3}\times \nb \\
& &[-4 M'' M^2 r^3+2 M'' M r^4+6 M'^2 M r^3-5  M'^2 r^4+\nb \\
& &8 M' \Lambda_g M^3 r^3+4 M' \Lambda_g M^2 r^4-8 M' M^3 r-\nb \\
& &14 M' M^2 r^2+11 M' M r^3-8 \dot M \Lambda_g M^3 r^3+\nb \\
& &8 \dot M M^3 r-2 \dot M M^2 r^2+24 \Lambda_g M^4 r^2-\nb \\
& &20 \Lambda_g M^3 r^3+8 M^4 +8 M^3 r-6 M^2 r^2]
\lb{Jphi}
\eqn

From the dynamical equation (\ref{dyn}) we have
\bq
D^{rr}=8 \pi \tau^{00}=\frac{d^{rr}}{128 \sqrt{r/M} M^{4} r}, 
\eq
where
\bqn
& &d^{rr}=\nb \\
& &-16 A' M^2 r^2- \sqrt{r/M} M'^2 \beta_0 r^3+\nb \\
& &2 \sqrt{r/M} M' \beta_0 M r^2-32 \sqrt{r/M} M' M^3 +\nb \\
& &96 \sqrt{r/M} \dot M M^3 -16 \sqrt{r/M} \dot M M^2 r-\nb \\
& &\sqrt{r/M} \beta_0 M^2 r-32 A \Lambda_g M^3 r^2+\nb \\
& &32 A M^3 -8 A M^2 r
\eqn
\bq
D^{\theta\theta}=8 \pi \tau^{\theta\theta}=\frac{d^{\theta\theta}}{32 \sqrt{r/M} M^3 r^3}, 
\eq
where
\bqn
& &d^{\theta\theta}=\nb \\
& &-8 A'' M^2 r^2+4 A' M' M r^2-12 A' M^2 r+\nb \\
& &32 \sqrt{r/M} \dot M' M^3 -16 \sqrt{r/M} M'' M^3+\nb \\
& &\sqrt{r/M} M'^2 \beta_0 r^2-2 \sqrt{r/M} M' \beta_0 M r+\nb \\
& &4 M' A M r-16 \sqrt{r/M} \ddot M M^3 +\nb \\
& &\sqrt{r/M} \beta_0 M^2 -32 A \Lambda_g M^3 r-4 A M^2
\eqn
\bq
D^{\phi\phi}=8 \pi \tau^{\phi\phi}=\frac{D^{\theta\theta}}{\sin^2 \theta}. 
\eq

\section{Possible Solutions}
We are looking for a HLT solution which is equivalent to the Vaidya's solution
in GRT. Then, initially we suppose that there is not any coupling between
the matter field (the null radiation) and the gauge field A. It means that
$J_A=0$. From equations (\ref{JA}) and
(\ref{Riccscalar}), we have
\bq
J_A=M'r^2-4 \Lambda_g M^2 r^2+4 M^2-2 M r=0,
\lb{JA1}
\eq
which give us the solution
\bq
M = 3 \frac{r^2}{-4 \Lambda_g r^3+12 r+3 f(t)},
\lb{Mrt}
\eq
where $f(t)$ is an integration time function.

Using equations (\ref{jdi}), (\ref{jdr}) and (\ref{jui}) we have
\bq
\frac{\dot M}{Mr\sqrt{r/M}}\left(1-\frac{M}{r}\right)=8\pi
\frac{1}{r^2}\frac{\dot M + M'}{\sqrt{r/M}}.
\lb{jdijdr}
\eq
Substituting equation(\ref{M*}) into (\ref{jdijdr}) we get
\bq
M(r,t)= \frac{r}{16\pi +1}.
\eq
So, we can see that the mass depends only on the coordinate $r$, i.e., it is
a static solution.  Therefore, we can conclude that there is not Vaidya's 
solution
in the theory of Ho\v{r}ava-Lifshitz, at least without coupling between the
null radiation and the gauge field A.

Moreover, since $M=M(V)$ and $U=t-r$, we can
write the equation (\ref{JA1}) in terms of $V$ and $U$, that is
\bq
\frac{1}{8}(V-U)^2 \frac{dM}{dV} - \Lambda_g M^2(V-U)^2 + 4M^2 -M(V-U) = 0.
\lb{JAUV1}
\eq
Deriving (\ref{JAUV1}) twice, in terms of U, and solving the differential
equation, we find
\bq
M(V)=-\frac{1}{M_0+8\Lambda_g V}.
\lb{MV}
\eq
Substituting the equation (\ref{MV}) into equation (\ref{JA1}), we find
that it does not satisfy for any $M_0$ and $\Lambda_g$.
Thus, again, we can conclude that, in fact,  there is no Vaidya's solution in 
the Ho\v{r}ava-Lifshitz theory if $J_A=0$.  In another words, the gauge field
$A$ must depend on the Vaidya's mass, i.e., $J_A=J_A(M)$.

Finally, let us suppose that $J_A\ne 0$, and considering the high complexity of 
the field equations, we use the follow ansatz $J_{\varphi}=0$ and a solution 
like $M(V)=M_0 V$, where
$M_0$ is a constant. We can show using equation (\ref{Jphi}) that we have no
consistent solution for it. This can imply in the need of the coupling
between the null radiation and the pre-potential $\varphi$ or, more
probably, that the particular solution proposed is not consistent.

\section{Conclusion}

In this paper, we have analyzed nonstationary radiative spherically symmetric 
spacetime, in general covariant theory of Ho\v{r}ava-Lifshitz 
gravity without the projectability condition and in the infrared limit. 
The Newtonian prepotential $\varphi$ was assumed null.
We have shown that the gauge field $A$ must interacts with the null radiation 
field of the Vaidya's spacetime in the HLT, since
we must have $J_A\ne 0$.
Besides, we can conclude that there is not Vaidya's solution, as we know in
GRT, in the theory of Ho\v{r}ava-Lifshitz.

\begin{acknowledgements}
The financial assistance from FAPERJ/UERJ (MFAdaS) are gratefully acknowledged.
The author (RC) acknowledges the financial support from FAPERJ 
(no. E-26/171.754/2000, E-26/171.533/2002, E-26/170.951/2006, E-26/110.432\-/2009 
and E26/111.714/2010). The authors (RC and MFAdaS) also acknowledge the 
financial support from Conselho Nacional de Desenvolvimento Cient\'ifico e
Tecnol\'ogico - CNPq - Brazil (no. 450572/2009-9, 301973/2009-1 and 
477268\-/2010-2). The author (MFAdaS) also acknowledges the financial support 
from Financiadora de Estudos e Projetos - FINEP - Brazil (Ref. 2399/03).
We also would like to thank Dr. Anzhong Wang for helpful discussions and
comments about this work.
\end{acknowledgements}

\section{Appendix A:  HLT Field Equations and Conservation Laws}

The variations of the action $S$ (\ref{TA}) with respect to $N$ and $N^{i}$ 
give rise to the Hamiltonian and momentum constraints,
\bqn \label{hami}
{\cal{L}}_K + {\cal{L}}_V^R + F_V-F_\varphi-F_\lambda= 8 \pi G J^t,\;\;
\eqn
\bqn \label{mom}
&&\nabla_j \bigg\{\pi^{ij} - \varphi {\cal{G}}^{ij} - \hat{{\cal{G}}}^{ijkl} a_l \nabla_k \varphi\nb\\
&&-(1-\lambda)g^{ij}\big(\nabla^2\varphi+a_k\nabla^k\varphi\big)\bigg\}=8\pi G J^i,
\lb{jui}
\eqn
where
\bqn
{\cal{L}}_V^R&=&\gamma_0 \zeta^2-R+\frac{\gamma_2 R^2+\gamma_3 R_{ij}R^{ij}}{\zeta^2}+\frac{\gamma_5}{\zeta^4} C_{ij}C^{ij},\nb\\
J^i &=& -N \frac{\delta {\cal{L}}_M}{\delta N_i},\;\;
J^t = 2 \frac{\delta (N {\cal{L}}_M)}{\delta N},\nb\\
\pi^{ij}&=&-K^{ij}+\lambda K g^{ij},
\eqn
and $F_V,\;F_\varphi$ and $F_\lambda$ are given, respectively, by
\bqn\label{a1}
F_V &=&  \beta_0 ( 2 a_i^i + a_i a^i) - \frac{\beta_1}{\zeta^2} \Bigg[3 (a_i a^i)^2 + 4 \nabla_i (a_k a^k a^i)\Bigg]\nb\\
    &&  +\frac{\beta_2}{\zeta^2}\Bigg[ (a_i^i)^2 + \frac{2}{N} \nabla^2 (N a_k^k)\Bigg]\nb\\
    && + \frac{\beta_3}{\zeta^2}\Bigg[ - (a_i a^i) a_j^j - 2 \nabla_i (a_j^j a^i) + \frac{1}{N} \nabla^2 (N a_i a^i)\Bigg]\nb\\
    &&+ \frac{\beta_4}{\zeta^2}\Bigg[ a_{ij} a^{ij} + \frac{2}{N} \nabla_j \nabla_i (N a^{ij})\Bigg]\nb\\
      && + \frac{\beta_5}{\zeta^2}\Bigg[- R (a_i a^i) - 2 \nabla_i (R a^i)\Bigg]\nb\\
      &&+ \frac{\beta_6}{\zeta^2}\Bigg[- a_i a_j R^{ij} - \nabla_i (a_j R^{ij})-\nabla_j (a_i R^{ij})\Bigg]\nb\\
      && +  \frac{\beta_7}{\zeta^2}\Bigg[ R a^i_i + \frac{1}{N} \nabla^2 (NR)\Bigg]\nb\\
      &&+ \frac{\beta_8}{\zeta^4}\Bigg[(\Delta a^i)^2 - \frac{2}{N} \nabla^i [\Delta (N \Delta a_i)]\Bigg],
\eqn
\bqn\label{a2}
F_\varphi &=& -  {\cal{G}}^{ij}\nabla_i \varphi \nabla_j \varphi, - \frac{2}{N} \hat{{\cal{G}}}^{ijkl} \nabla_l (N K_{ij} \nabla_k \varphi),\nb\\
        &&  - \frac{4}{3}\Bigg[ \hat{{\cal{G}}}^{ijkl} \nabla_l (\nabla_k \varphi \nabla_i \nabla_j \varphi)\Bigg]\nb\\
         &&+ \frac{5}{3}\Bigg[ -  \hat{{\cal{G}}}^{ijkl} [(a_i \nabla_j \varphi) (a_k \nabla_l \varphi)+\nabla_i(a_k\nabla_j\varphi\nabla_l\varphi)\nb\\
         && +\nabla_k(a_i\nabla_j\varphi\nabla_l\varphi)]\Bigg]\nb\\
         &&+ \frac{2}{3}\Bigg[ \hat{{\cal{G}}}^{ijkl}[a_{ik} \nabla_j \varphi \nabla_l \varphi + \frac{1}{N} \nabla_i\nabla_k (N\nabla_j\varphi\nabla_l\varphi)]\Bigg],\nb\\
\eqn
\bqn\lb{a3a}
F_\lambda &=& (1-\lambda) \Bigg\{(\nabla^2 \varphi + a_i \nabla^i \varphi)^2 - \frac{2}{N} \nabla_i (NK\nabla^i \varphi)\nb\\
           && - \frac{2}{N} \nabla_i [N (\nabla^2 \varphi + a_i \nabla^i \varphi) \nabla^i \varphi]\Bigg\}\label{a3}.
\eqn

$\left(F_n\right)_{ij}$, $\left(F^{a}_{s}\right)_{ij}$ and $\left(F^{\varphi}_{q}\right)_{ij}$, defined in equation (\ref{tauij}),  are given, respectively, by
\bqn
(F_0)_{ij} &=& -\frac{1}{2}g_{ij},\nb\\
(F_1)_{ij} &=& R_{ij}-\frac{1}{2}Rg_{ij}+\frac{1}{N}(g_{ij}\nabla^2 N-\nabla_j\nabla_i N),\nb\\
(F_2)_{ij} &=& -\frac{1}{2}g_{ij}R^2+2RR_{ij}\nb\\
             &&  +\frac{2}{N}\left[g_{ij}\nabla^2(NR)-\nabla_j\nabla_i(NR)\right],\nb\\
(F_3)_{ij} &=& -\frac{1}{2}g_{ij}R_{mn}R^{mn}+2R_{ik}R^k_{j}\nb\\
             &&  +\frac{1}{N}\Big[- 2\nabla_k\nabla_{(i}(NR_{j)}^k)\nb\\
             &&  +\nabla^2(NR_{ij})+g_{ij}\nabla_m\nabla_n(NR^{mn})\Big], \nb\\
(F_4)_{ij} &=&  -\frac{1}{2}g_{ij}R^3+3R^2R_{ij}\nb\\
             &&   +\frac{3}{N}\Big(g_{ij}\nabla^2-\nabla_j\nabla_i\Big)(NR^2),\nb\\
(F_5)_{ij} &=& -\frac{1}{2}g_{ij}RR_{mn}R^{mn}\nb\\
             &&+R_{ij}R_{mn}R^{mn}+2RR_{ik}R^k_{j}\nb\\
             &&  +\frac{1}{N}\Big[g_{ij}\nabla^2(NR_{mn}R^{mn})\nb\\
             &&-\nabla_j\nabla_i(NR_{mn}R^{mn})\nb\\
             &&  +\nabla^2(NRR_{ij})+g_{ij}\nabla_m\nabla_n(NRR^{mn})\nb\\
             &&  -2\nabla_m\nabla_{(i}(R^m_{j)}NR)\Big], \nb\\
(F_6)_{ij} &=& -\frac{1}{2}g_{ij}R^m_nR^n_lR^l_m+3R^{mn}R_{mi}R_{nj}\nb\\
             &&  +\frac{3}{2N}\Big[g_{ij}\nabla_m\nabla_n(NR^m_aR^{na}) \nb\\
             &&+ \nabla^2(NR_{mi}R^m_j)
              -2\nabla_m\nabla_{(i}(NR_{j)n}R^{mn})\Big], \nb\\
(F_7)_{ij} &=& -\frac{1}{2}g_{ij}R\nabla^2R+R_{ij}\nabla^2R+R\nabla_i\nabla_j R\nb\\
             &&  +\frac{1}{N}\Big[g_{ij}\nabla^2(N\nabla^2R)-\nabla_j\nabla_i(N\nabla^2R)\nb\\
             &&+R_{ij}\nabla^2(NR)
              +g_{ij}\nabla^4(NR)-\nabla_j\nabla_i (\nabla^2 (NR))\nb\\
             &&  - \nabla_{(j}(NR\nabla_{i)}R)+\frac{1}{2}g_{ij}\nabla_k(NR\nabla^kR)\Big], \nb\\
(F_8)_{ij} &=& -\frac{1}{2}g_{ij}(\nabla_mR_{nl})^2 + 2 \nabla^mR^n_i\nabla_mR_{nj}\nb\\
             &&  +\nabla_iR^{mn}\nabla_jR_{mn}+\frac{1}{N}\Big[2 \nabla_n\nabla_{(i}\nabla_m(N\nabla^mR^n_{j)})\nb\\
             &&  -\nabla^2\nabla_m(N\nabla^mR_{ij})-g_{ij}\nabla_n\nabla_p\nabla_m(N\nabla^mR^{np})\nb\\
             &&  -2\nabla_m(NR_{l(i}\nabla^mR^l_{j)})-2\nabla_n(NR_{l(i}\nabla_{j)}R^{nl})\nb\\
             &&  +2\nabla_k(NR^k_l\nabla_{(i}R^l_{j)})\Big], \nb\\
(F_9)_{ij} &=& -\frac{1}{2} g_{ij} a_k G^k+\frac{1}{2} \Big[a^k R_{k(j} \nabla_{i)} R + a_{(i} R_{j)k} \nabla^k R\Big]\nb\\
           &&-a_kR_{mi}\nabla_jR^{mk}-a^kR_{n(i}\nabla^nR_{j)k}\nb \\
           &&-\frac{1}{2}\Big[a_iR^{km}\nabla_mR_{kj}+a_jR^{km}\nabla_mR_{ki}\Big]\nb\\
           &&-\frac{3}{8}a_{(i}R\nabla_{j)}R+\frac{3}{8}\Bigg\{R\nabla_k(Na^k)R_{ij}\nb\\
           &&+g_{ij}\nabla^2\Big[R\nabla_k(Na^k)\Big]-\nabla_i\nabla_j\Big[R\nabla_k(Na^k)\Big]\Bigg\}\nb\\
           &&+\frac{1}{4N} \Bigg\{- \frac{1}{2}\nabla^m \Big[\nabla_{(i} Na_{j)}\nabla_m R+\nabla_{(i}(\nabla_{j)}R) Na_m\Big]\nb\\
           &&+\nabla^2 (N a_{(i}\nabla_{j)}R)+g_{ij} \nabla^m\nabla^n (Na_m\nabla_nR)\nb\\
           && +\nabla^m\Big[\nabla_{(i} (\nabla_{j)} R^k_m) Na_k+\nabla_{(i}(\nabla_m R^k_{j)})Na_k\Big]\nb\\
           &&-2\nabla^2(Na_k\nabla_{(i} R^k_{j)})-2g_{ij}\nabla^m\nabla^n(Na_k\nabla_{(n}R_{m)}^k)\nb\\
           &&- \nabla^m \Big[\nabla_i\nabla_p(Na_jR_m^p+Na_mR_j^p)\nb\\
           &&+\nabla_j\nabla_p(Na_iR_m^p+Na_mR_i^p)\Big]\nb\\
           &&+2\nabla^2\nabla_p(Na_{(i}R_{j)}^p)\nb\\
           && +2g_{ij}\nabla^m\nabla^n \nabla^p(Na_{(n}R_{m)p})\Bigg\}, \nb\\
\eqn

\bqn
(F_0^a)_{ij} &=&  -\frac{1}{2} g_{ij} a^k a_k +a_i a_j, \nb\\
(F_1^a)_{ij} &=&  -\frac{1}{2} g_{ij} (a_k a^k)^2 + 2 (a_k a^k) a_i a_j,\nb\\
(F_2^a)_{ij} &=&  -\frac{1}{2} g_{ij} (a_k^k)^2 + 2 a_k^k a_{ij}\nb\\
             &&   - \frac{1}{N} \Big[2 \nabla_{(i} (N a_{j)} a_k^k) - g_{ij} \nabla_\alpha (a_\alpha N a_k^k)\Big],\nb\\
(F_3^a)_{ij} &=&   -\frac{1}{2} g_{ij} (a_k a^k) a_\beta^\beta + a^k_k a_ia_j + a_k a^k a_{ij}\nb\\
             &&   - \frac{1}{N} \Big[ \nabla_{(i} (N a_{j)} a_k a^k) - \frac{1}{2} g_{ij} \nabla_\alpha (a_\alpha N a_ka^k)\Big],\nb\\
(F_4^a)_{ij} &=&  - \frac{1}{2} g_{ij} a^{mn} a_{mn} + 2a^k_i a_{kj} \nb\\
             &&   - \frac{1}{N} \Big[\nabla^k (2 N a_{(i} a_{j)k} - N a_{ij} a_k)\Big], \nb\\
(F_5^a)_{ij} &=&  -\frac{1}{2} g_{ij} (a_k a^k ) R + a_i a_j R + a^k a_k R_{ij} \nb\\
              &&  + \frac{1}{N} \Big[ g_{ij} \nabla^2 (N a_k a^k) - \nabla_i \nabla_j (N a_k a^k)\Big], \nb\\ 
(F_6^a)_{ij} &=&   -\frac{1}{2} g_{ij} a_m a_n R^{mn} +2 a^m R_{m (i} a_{j)} \nb\\
              &&  - \frac{1}{2N} \Big[ 2 \nabla^k \nabla_{(i} (a_{j)} N a_k) - \nabla^2 (N a_i a_j) \nb\\
               && - g_{ij} \nabla^m \nabla^n (N a_m a_n)\Big], \nb\\ 
(F_7^a)_{ij} &=&  -\frac{1}{2} g_{ij} R a_k^k + a_k^kR_{ij} + R a_{ij} \nb\\
             &&   + \frac{1}{N} \Big[ g_{ij} \nabla^2 (N a_k^k) - \nabla_i \nabla_j (N a_k^k) \nb\\
             &&  - \nabla_{(i} (N R a_{j)}) + \frac{1}{2} g_{ij} \nabla^k (N R a_k)\Big], \nb\\
(F_8^a)_{ij} &=&  -\frac{1}{2} g_{ij} (\Delta a_k)^2 + (\Delta a_i) (\Delta a_j) + 2 \Delta a^k \nabla_{(i} \nabla_{j)} a_k \nb\\
             &&   + \frac{1}{N} \Big[\nabla_k [a_{(i} \nabla^k (N \Delta a_{j)}) + a_{(i} \nabla_{j)} (N \Delta a^k)\nb\\
             &&   - a^k \nabla_{(i} (N \Delta a_{j)}) + g_{ij} N a^{\beta k} \Delta a_\beta  - N a_{ij} \Delta a^k ]\nb\\
             &&  -  2 \nabla_{(i} (N a_{j)k} \Delta a^k)\Big],
\eqn

\bqn
(F_1^\varphi)_{ij} &=&     -\frac{1}{2} g_{ij} \varphi {\cal{G}}^{mn}K_{mn}\nb\\
                   &&  + \frac{1}{2\sqrt{g} N}  \partial_t (\sqrt{g} \varphi {\cal{G}}_{ij}) -
                                                                                2 \varphi K_{(i}^\nu R_{j) \nu} \nb\\
                   &&       + \frac{1}{2} \varphi (K R_{ij} + K_{ij} R - 2 K_{ij} \Lambda_g ) \nb\\
                   &&      + \frac{1}{2N} \bigg\{{\cal{G}}_{ij} \nabla^k (\varphi N_k) - 2 {\cal{G}}_{k(i} \nabla^k (N_{j)} \varphi)\nb\\
                   &&      +  g_{ij}\nabla^2 (N \varphi K) - \nabla_i \nabla_j (N \varphi K) \nb\\
                   &&+ 2  \nabla^k \nabla_{(i} (K_{j)k} \varphi N),\nb\\
                   &&      - \nabla^2 (N \varphi K_{ij}) - g_{ij} \nabla^\alpha \nabla^\beta (N \varphi K_{\alpha \beta})\bigg\}, \nb\\
(F_2^\varphi)_{ij} &=&   - \frac{1}{2}g_{ij} \varphi {\cal{G}}^{mn} \nabla_m\nabla_n \varphi \nb\\
                   &&     - 2 \varphi \nabla_{(i} \nabla^k R_{j)k} + \frac{1}{2} \varphi (R- 2 \Lambda_g) \nabla_i \nabla_j \varphi \nb\\
                   &&     - \frac{1}{N} \bigg\{- \frac{1}{2} (R_{ij} + g_{ij} \nabla^2 - \nabla_i \nabla_j )(N \varphi \nabla^2 \varphi)\nb\\
                   &&      - \nabla_k \nabla_{(i} (N \varphi \nabla^k \nabla_{j)} \varphi) + \frac{1}{2}\nabla^2 (N \varphi \nabla_i \nabla_j \varphi) \nb\\
                   &&      + \frac{g_{ij}}{2} \nabla^\alpha \nabla^\beta ( N \varphi \nabla_\alpha \nabla_\beta \varphi)\nb\\
                   &&      - {\cal{G}}_{k (i} \nabla^k (N \varphi \nabla_{j)}\varphi ) + \frac{1}{2} {\cal{G}}_{ij} \nabla^k (N \varphi \nabla_k \varphi)\bigg\},\nb\\
(F_3^\varphi)_{ij} &=&    - \frac{1}{2}g_{ij} \varphi {\cal{G}}^{mn} a_m\nabla_n \varphi \nb\\
                   &&    -\varphi ( a_{(i} R_{j)k} \nabla^k \varphi + a^k R_{k(i} \nabla_{j)} \varphi)\nb\\
                   &&      + \frac{1}{2} (R - 2 \Lambda_g)  \varphi a_{(i} \nabla_{j)} \varphi \nb\\
                   &&      - \frac{1}{N}\bigg\{ - \frac{1}{2} (R_{ij} + g_{ij} \nabla^2 - \nabla_i \nabla_j ) (N \varphi a^k \nabla_k \varphi)\nb\\
                   &&     - \frac{1}{2} \nabla^k \Big[  \nabla_{(i} (\nabla_{j)} \varphi N \varphi)+ \nabla_{(i} (a_{j)} \varphi N \nabla_k \varphi) \Big] \nb\\
                   &&     + \frac{1}{2}\nabla^2 (N \varphi a_{(i} \nabla_{j)} \varphi) \nb\\
                   &&+ \frac{g_{ij}}{2} \nabla^\alpha \nabla^\beta (N \varphi a_\alpha \nabla_\beta \varphi)\bigg\}, \nb\\ 
(F_4^\varphi)_{ij} &=&    - \frac{1}{2}g_{ij} \hat{{\cal{G}}}^{mnkl}K_{mn} a_{(k}\nabla_{l)}\varphi \nb\\
                   &&    + \frac{1}{2 \sqrt{g} N} \partial_t [\sqrt{g} {\cal{G}}_{ij}^{\;\;k l} a_{(l} \nabla_{k)} \varphi] \nb\\
                   &&      + \frac{1}{2N} \nabla^\alpha \Big[  a_\alpha N_{(i} \nabla_{j)} \varphi +  N_{(i} a_{j)} \nabla_\alpha  \varphi \nb\\
                   &&      -  N_\alpha a_{(i} \nabla_{j)} \varphi + 2 g_{ij} N_\alpha a^k \nabla_k \varphi \Big] \nb\\
                   &&      + \frac{1}{N} \nabla_{(i} (N N_{j)} a^k \nabla_k \varphi)\nb\\
                   &&      + a^k K_{k(i} \nabla_{j)} \varphi + a_{(i} K_{j)k} \nabla^k \varphi \nb\\
                   &&      - K a_{(i} \nabla_{j)} \varphi - K_{ij} a^k \nabla_k \varphi, \nb\\
(F_5^\varphi)_{ij} &=&     -\frac{1}{2} g_{ij} \hat{{\cal{G}}}^{mnkl}[a_{(k}\nabla_{l)}\varphi][\nabla_m\nabla_n\varphi]\nb\\
                    && -a_{(i} \nabla^k \nabla_{j)} \varphi \nabla_k \varphi - a_k \nabla^k \nabla_{(i} \varphi \nabla_{j)}\varphi\nb\\
                   &&     + a_{(i} \nabla_{j)} \varphi \nabla^2 \varphi + a^k \nabla_k \varphi \nabla_i\nabla_j \varphi \nb\\
                   &&     + \frac{1}{2N} \bigg\{ \nabla^k (N \varphi a_k \nabla_i \varphi \nabla_j \varphi) \nb\\
                   &&   - 2 \nabla_{(i} (N \nabla_{j)} \varphi a^k
                          \nabla_k \varphi) \nb\\
                   &&     +g_{ij} \nabla^\alpha (\nabla_\alpha \varphi a^k \nabla_k \varphi)\bigg\}, \nb\\
(F_6^\varphi)_{ij} &=&   - \frac{1}{2} g_{ij}\hat{{\cal{G}}}^{mnkl} [a_{(m}\nabla_{n)}\varphi][a_{(k}\nabla_{l)}\varphi]\nb\\
                     &&-\frac{1}{2} (a^k \nabla_i \varphi- a_i \nabla^k \varphi) (a_k \nabla_j \varphi - a_j \nabla_k \varphi), \nb\\
(F_7^\varphi)_{ij} &=&   -\frac{1}{2} g_{ij}   \hat{{\cal{G}}}^{mnkl} [\nabla_{(n}\varphi][a_{m)(k}][\nabla_{l)}\varphi]\nb\\
                    && -\frac{1}{2} a_k^k \nabla_i \varphi \nabla_j \varphi - \frac{1}{2} a_{ij} \nabla^k \varphi \nabla_k \varphi \nb\\
                     &&     +  a^k_{(i} \nabla_{j)} \varphi \nabla_k \varphi - \frac{1}{2N}\bigg \{- \nabla_{(i} (N a_{j)} \nabla_k \varphi \nabla^k \varphi) \nb\\ &&+ \nabla^k (N a_{(i} \nabla_{j)} \varphi \nabla_k \varphi)\nb\\
                   &&      + \frac{g_{ij}}{2} \nabla^k (N a_k \nabla^m \varphi \nabla_m \varphi) \nb\\
                   &&- \frac{1}{2}\nabla^k (N a_k \nabla_i \varphi \nabla_j \varphi)\bigg\}, \nb\\
(F_8^\varphi)_{ij} &=&    - \frac{1}{2} g_{ij} (\nabla^2 \varphi+a_k\nabla^k\varphi)^2\nb\\
                   &&    -2 (\nabla^2 \varphi + a_k \nabla^k \varphi) (\nabla_i \nabla_j \varphi + a_i \nabla_j \varphi ) \nb\\
                   &&      -\frac{1}{N} \bigg \{ - 2 \nabla_{(j} [N \nabla_{i)} \varphi (\nabla^2 \varphi + a_k \nabla^k \varphi)] \nb\\
                   &&       + g_{ij} \nabla^\alpha [N (\nabla^2 \varphi + a_k \nabla^k \varphi) \nabla_\alpha \varphi]\bigg\}, \nb\\
(F_9^\varphi)_{ij} &=&    - \frac{1}{2} g_{ij}(\nabla^2 \varphi+a_k\nabla^k\varphi)K \nb\\
                    && - (\nabla^2 \varphi + a_k \nabla^k \varphi) K_{ij} \nb\\
                   &&     - (\nabla_i \nabla_j \varphi + a_i \nabla_j \varphi ) K \nb\\
                  &&     +\frac{1}{2 \sqrt{g} N}  \partial_t [\sqrt{g} (\nabla^2 \varphi + a_k \nabla^k \varphi) g_{ij}] \nb\\
                    &&      - \frac{1}{N}\bigg\{ - \nabla_{(j} [ N_{i)} (\nabla^2 \varphi + a_k \nabla^k \varphi)] \nb\\
                    &&     + \frac{1}{2} g_{ij} \nabla_\alpha [ N_\alpha (\nabla^2 \varphi + a_k \nabla^k \varphi) ]\nb\\
                    &&    -  \nabla_{(j} (N K \nabla_{i)} \varphi) + \frac{1}{2} g_{ij} \nabla_k (N K \nabla^k \varphi)
                            \bigg\}.\nb\\
\eqn

Variations of $S$ with respect to $\varphi$ and $A$ yield, respectively,
\bqn \label{phi}
&& \frac{1}{2} {\cal{G}}^{ij} ( 2K_{ij} + \nabla_i\nabla_j\varphi  +a_{(i}\nabla_{j)}\varphi)\nb\\
&& + \frac{1}{2N} \bigg\{ {\cal{G}}^{ij} \nabla_j\nabla_i(N \varphi) - {\cal{G}}^{ij} \nabla_j ( N \varphi a_i)\bigg\}\nb\\
&& - \frac{1}{N} \hat{{\cal{G}}}^{ijkl} \bigg \{ \nabla_{(k} ( a_{l)} N K_{ij}) + \frac{2}{3} \nabla_{(k} (a_{l)} N \nabla_i \nabla_j \varphi)\nb\\
&& - \frac{2}{3} \nabla_{(j} \nabla_{i)} (N a_{(l} \nabla_{k)} \varphi) + \frac{5}{3} \nabla_j (N a_i a_k \nabla_l \varphi)\nb\\
&& + \frac{2}{3} \nabla_j (N a_{ik} \nabla_l \varphi)\bigg\} \nb\\
&& + \frac{1-\lambda}{N} \bigg\{\nabla^2  \left.[N (\nabla^2 \varphi + a_k \nabla^k \varphi)\right.] \nb\\
&& - \nabla^i [N(\nabla^2 \varphi + a_k \nabla^k \varphi) a_i] \nb\\
&&+ \nabla^2 (N K) - \nabla^i ( N K a_i)\bigg \}
 = 8 \pi G J_\varphi,
\eqn
and
\bqn \label{JA}
R-2 \Lambda_g = 8 \pi G J_A,
\eqn
where
\bqn
J_\varphi = -\frac{\delta {\cal{L}}_M}{\delta \varphi},\;\;\;
 J_A= 2 \frac{\delta ( N {\cal{L}}_M)}{\delta A}.
\eqn

On the other hand, the variation of $S$ with respect to $g_{ij}$ yields the 
dynamical equations,
\bqn \label{dyn}
\frac{1}{\sqrt{g}N} \frac{\partial}{\partial t}\left(\sqrt{g} \pi^{ij}\right)+2(K^{ik}K^j_k-\lambda K K^{ij})\nb\\
-\frac{1}{2}g^{ij}{\cal{L}}_K+\frac{1}{N}\nabla_k (\pi^{ik}N^j+\pi^{kj}N^i-\pi^{ij}N^k)\nb\\
+F^{ij}+F^{ij}_a-\frac{1}{2}g^{ij}{\cal{L}}_A+F^{ij}_\varphi\nb\\
-\frac{1}{N}(AR^{ij}+g^{ij}\nabla^2A-\nabla^j\nabla^iA)
=8\pi G \tau^{ij},\;\;\;\;\;\;
\eqn
where
\bqn
\lb{tauij}
\tau^{ij}&=&\frac{2}{\sqrt{g}N} \frac{\delta(\sqrt{g}N{\cal{L}}_M)}{\delta g_{ij}}, \nb\\
F^{ij}&=&\frac{1}{\sqrt{g}N}\frac{\delta (-\sqrt{g}N {\cal{L}}_V^R)}{\delta g_{ij}}\nb\\
&=& \sum_{s=0}\hat{\gamma}_s\zeta^{n_s}(F_s)^{ij},\;\;\;\\
F^{ij}_a&=&\frac{1}{\sqrt{g}N}\frac{\delta (-\sqrt{g}N {\cal{L}}_V^a)}{\delta g_{ij}}\nb\\
& =& \sum_{s=0}\beta_s\zeta^{m_s}(F_s^a)^{ij},\\
F^{ij}_\varphi&=&\frac{1}{\sqrt{g}N}\frac{\delta (-\sqrt{g}N {\cal{L}}_V^\varphi)}{\delta g_{ij}}\nb\\
&=& \sum_{s=0}\mu_s(F_s^\varphi)^{ij},
\eqn
with
\bqn
\hat{\gamma}_s &=& \left(\gamma_0, \gamma_1, \gamma_2, \gamma_3, \frac{1}{2}\gamma_5, -\frac{5}{2}\gamma_5, 3\gamma_5, \frac{3}{8}\gamma_5, \gamma_5, \frac{1}{2}\gamma_5\right), \nb\\
n_s &=& (2, 0, -2, -2, -4, -4, -4, -4, -4, -4),\nb\\
m_s&=& (0, -2,-2,-2, -2, -2, -2, -2, -4 ), \nb\\
\mu_s &=& \left(2, 1, 1, 2, \frac{4}{3}, \frac{5}{3}, \frac{2}{3}, 1-\lambda, 2-2 \lambda\right).
\eqn

In addition, the matter components $(J^t, J^i, J_\varphi, J_A, \tau^{ij})$ 
satisfy the conservation laws of energy and momentum,
\bqn
\label{energy conservation}
\int d^3x \sqrt{g} N \bigg[\dot{g}_{ij}\tau^{ij}-\frac{1}{\sqrt{g}}\partial_t (\sqrt{g} J^t)+\frac{2 N_i}{\sqrt{g} N}\partial_t (\sqrt{g} J^i)\nb\\
-\frac{A}{\sqrt{g} N}\partial_t (\sqrt{g} J_A)-2\dot{\varphi}J_\varphi\bigg]=0,\;\;\;\;\;\;\;\;\\
\label{mom conservation}
\frac{1}{N}\nabla^i(N\tau_{ik})-\frac{1}{\sqrt{g} N} \partial_t (\sqrt{g} J_k)-\frac{J_A}{2N}\nabla_k A-\frac{J^t}{2N}\nabla_kN\nb\\
-\frac{N_k}{N} \nabla_i J^i-\frac{J_i}{N}(\nabla_i N_k-\nabla_k N_i)+J_\varphi \nabla_k\varphi=0.\;\;\;\;\;\;\;\;
\eqn

\end{document}